# Observation of a surface lattice resonance in a fractal arrangement of gold nanoparticles

Ting Lee Chen[1], Jord C. Prangsma[1,2], Frans B. Segerink[1], Dirk Jan Dikken[1], and Jennifer L. Herek[1]*

[1]Optical Sciences Group, MESA+ Institute for Nanotechnology, University of Twente, 7500 AE Enschede, The Netherlands
[2]Nanobiophysics, MESA+ Institute for Nanotechnology, University of Twente, 7500 AE Enschede, The Netherlands

*Corresponding author: j.l.herek@utwente.nl

**The collective response of closely spaced metal particles in non-periodic arrangements has the potential to provide a beneficial angular and frequency dependence in sensing applications. In this paper, we investigate the optical response of a Sierpinski fractal arrangement of gold nanoparticles and show that it supports a collective resonance similar to the surface lattice resonances that exist in periodic arrangements of plasmonic resonators. Using back focal plane microscopy, we observe the leakage of radiation out of a surface lattice resonance that is efficiently excited when the wavenumber of the incident light matches a strong Fourier component of the fractal structure. The efficient coupling between localized surface plasmons leads to a collective resonance and a Fano-like feature in the scattering spectrum. Our experimental observations are supported by numerical simulations based on the coupled-dipole approximation and finite-difference time-domain methods. This work presents a first step towards the application of fractal arrangements for plasmonic applications**

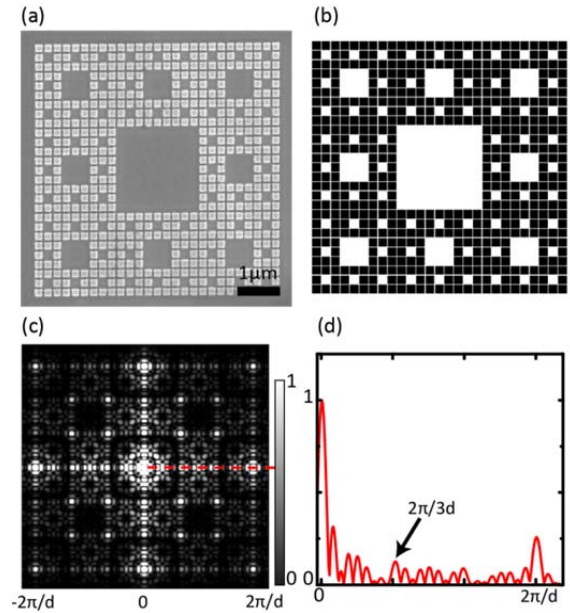

Fig. 1 (a) SEM image of the Sierpinski carpet carved out of gold on a glass substrate; scale bar 1 μm. (b) Idealized version of the structure layout. (c) 2D Fourier transform of the structure design in b, d is the center-to-center particle distance. The red line indicates the cross section along which Fig. 1(d) is shown. (d) The profile of the amplitude of the Fourier transform along the x-direction.

## 1. INTRODUCTION

Metal particles with subwavelength dimensions have resonances in their scattering and absorption spectra in the optical regime called localized plasmon resonances [1]. Associated with these resonances are strong electromagnetic field enhancements near the particle surface. The enhancement of light-matter interaction that this offers can be used in sensing or lighting technologies [2-4]. Single metal particles can be tuned in spectrum by simple geometrical modifications such as changes in their length-to-width ratio [5], however, single-particle plasmon resonances have a very broad spectral response due to their dipolar nature and associated efficient radiation losses. The dipolar nature also dictates limited control over their angular-dependent response. Isolated particles thus are limited in sensitivity of detection mechanisms based on changes in spectrum or angular response.

To take control over the spectral and angular response, arrangements of multiple metal particles of similar shape in close proximity can be used. Studies have shown how the coupling between the particles gives rise to many different phenomena such as mode splitting and Fano resonances in dimers, trimers, quadrimers and heptamers [6-10]. To make plasmonic resonances with a narrower linewidth, a reduction of the radiation loss is needed. Such a reduction can be achieved by constructing resonances of higher-order multipolar nature as these are less efficient in coupling to the far-field, so-called 'dark' or subradiant modes.

One particularly efficient way to create a subradiant resonance is to make large area periodic arrangements of particles. Several studies have shown how resonant interactions of localized plasmons in periodic arrangements lead to narrow linewidth spectral features [11-16]. The origin of these resonances was shown to be well described by a coupled-dipole model in which the broad localized surface plasmon resonances of the particles are described as dipolar resonators [15]. The coupling between the particles leads to sharp spectral features near the condition for the (dis)appearance of a diffraction order, i.e. when the reciprocal lattice constant $k_d=2\pi/d$ crosses the light line $k=\omega/c$ with $d$ the lattice spacing, $\omega$ the angular frequency and $c$ the speed of light in the medium. To describe the collective resonance of the particles we use the often-employed term surface lattice resonance.

Because the sharp linewidths arise from the collective in-phase oscillation of the monomers, these features are typically not only sharp in spectra but also sharp in their angular sensitivity. Since this property may be disadvantageous in applications, it is worthwhile to explore how similar behavior arises in non-periodic yet extended arrangements of particles with simple design rules such as quasi-periodic or fractal arrangements. Such configurations generate more densely-packed features in reciprocal space [17, 18] and, as a result, more isotropic behavior of diffracted light in terms of radiated directions or features that are less regularly spectrally spaced can be expected [18-20].

In this article, we study a fractal antenna structure based on the Sierpinski carpet geometry and visualize a surface lattice resonance using leakage radiation microscopy in the Fourier plane [21, 22]. The coupled-dipole approximation and finite-difference time-domain numerical simulations are adopted to validate and understand our findings. We will relate our findings with two different viewpoints on the structure. One is making a relation with surface lattice resonances in periodic particle arrays, showing that the Fourier amplitude of the fractal plays a similar role as the reciprocal lattice vector in gratings and regular arrays. We show that surface lattice resonances, that generate the dominant spectral features in periodic structures, also occur in fractal geometries. Secondly, we relate spectral features in the measurement with the most dominant eigenmode found by using an eigen-decomposition method. In the end, the appearance of Fano line shapes

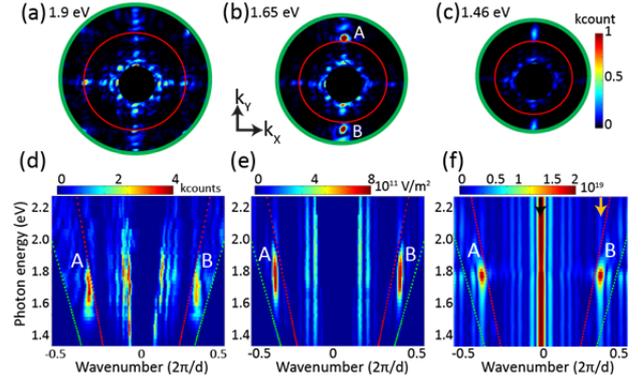

Fig. 2 Panels (a) - (c) show the BFP images of the Sierpinski carpet optical antenna with the incident light photon energy at 1.9, 1.65 and 1.46 eV, respectively. The CCD counts are normalized to the incident light power, exposure time and quantum efficiency of CCD. Note that the central region, where the 0$^{th}$ order diffracted light would appear in the image, is blocked by the beam blocker. (d) The cross section of the BFP image along the y-direction for different wavelengths. (e) The cross section of the far-fields angular radiation along the y-direction as obtained by FDTD. (f) The Fourier transform of the dipole moments obtained by coupled dipole approximation, $\tilde{P}_x(k_x, k_y)$. The black and orange arrow indicate correspond to the traces presented in Fig. 4(c). The red and green dashed lines in d,e,f indicate the light line in air and 1.4$k_0$ respectively.

[23, 7] in the scattering spectrum of the antenna is discussed.

Note that diffraction phenomena of fractal structures have been a topic of fundamental study since the 1970s [24-26]. More recently, optical fractal antennas have been proposed in which localized surface plasmon resonances of metal nanoparticles are to serve as an interface between propagative and localized light fields [27, 28]. Created from hundreds of metal nanoparticles arranged in a fractal pattern, fractal optical antennas have already been realized [29], but up to now only few studies have explored the possibilities that these structures offer. In principle the self-similarity on different length scales makes them ideally suited for frequency-independent design. As such, fractal antennas are currently used in broadband wireless communication in the radiofrequency region [30], such characteristics are however not considered in this article.

## 2. SAMPLE

The Sierpinski carpet morphology studied in this paper is constructed following the suggestion of Volpe et al. [27]. We start from a square monomer with size of 200 (width) × 200 (length) × 40 (thickness) nm$^3$ which is the 0$^{th}$ order structure. This monomer is copied into a 3×3 array and the central one is removed, creating the 1$^{st}$ order unit of the structure. The gap distance between each monomer is 20 nm. The same procedure is recursively applied to construct the next order of the fractal pattern. Fig. 1(a) shows a scanning electron micrograph of a 3$^{rd}$ order Sierpinski carpet as it is fabricated from a single crystalline gold flake by focused ion beam milling on a glass substrate [31] with a 100 nm ITO layer. The center-to-center spacing of the particles d is 220 nm. Note that while typically a Sierpinski carpet is a continuous construction with segmented holes or otherwise filled units, here we present what is essentially an inverse structure, with individual gold segments arranged in the Sierpinski carpet pattern and small gaps in between.

It is instructive to look at the Fourier transform of the idealized structure (Fig. 1b), as shown in Fig. 1c. It shows the magnitude of Fourier components as a function of the spatial frequency in x and y direction. While a periodic square lattice has a reciprocal lattice built up from the reciprocal lattice constant $k_d$ and higher harmonics as $\boldsymbol{G}_{per} = nk_d\boldsymbol{x} + mk_d\boldsymbol{y}$ with n and m integers, the fractal structure has many lower harmonics and can be written in general as:

$$\boldsymbol{G}_{frac} = \sum_i n_i\, k_d \left(\frac{1}{3}\right)^i \boldsymbol{x} + \sum_j m_j\, k_d \left(\frac{1}{3}\right)^j \boldsymbol{y},$$

where $i$ runs from 0 to the order of the fractal. The region in Fourier space within the "unit cell" $|\boldsymbol{k}| < \frac{2\pi}{d}$ of the Fourier transform of the Sierpinski lattice is therefore densely filled instead of empty as in a periodic structure.

### 3. BACK-FOCAL PLANE MICROSCOPY

Back-focal plane (BFP) microscopy is used to image the diffraction pattern of the optical antenna. In our setup, wavelength filtered light (from 550 – 950 nm) of a supercontinuum source is weakly focused on the sample from the air side. The incident polarization is aligned along the x direction. The forward-scattered diffracted light is collected using a 1.4 NA oil immersion objective, where the central part (including the directly transmitted beam and 0$^{th}$ order diffracted light) is blocked by a beam stop. An image of the BFP is made on a sensitive CCD camera after selecting x-polarized light. Because the incident wavelength is known, a measurement of the BFP is a direct measurement of the in-plane wavevector **k**=(k$_x$, k$_y$) of the collected light. More details of the setup are presented in the supplementary materials.

### 4. EXPERIMENTAL RESULTS

Figs. 2(a)-(c) show three diffraction patterns observed in the BFP of the microscope. These images show diffraction by the optical antenna upon illumination with light polarized along the x-direction with incident photon energy 1.9, 1.65 and 1.46 eV (650, 750 and 850 nm), respectively. The inner red circles indicate the wavenumber of the light line in air (k$_0$), whereas the green circle corresponds to the wavenumber that can be maximally collected with the 1.4 NA objective (1.4k$_0$). The measured diffraction patterns resemble the calculated Fourier amplitude of the Sierpinski carpet geometry shown in Fig. 1(c). As the incident photon energy decreases, the diffraction peaks gradually disappear beyond the observable region of the back-focal plane microscope. We note that the diffraction patterns also include spots corresponding to scattered light with wavenumbers larger than that of the incident wavenumber in air i.e. beyond the red circles in each panel of Fig. 2. These wavenumbers correspond to evanescent waves in air, which can be observed in the BFP image due to the 1.4 NA oil immersion objective in our microscope.

The most striking feature in the BFP images of Fig. 2 is the appearance of two intense spots at the incident photon energy of 1.65 eV marked A and B. The diffraction signal in this region of k-space extending beyond the light line in air is much weaker when incident light of 1.9 and 1.46 eV is used, suggesting a significant spectral dependence. To elucidate the spectral dependence of these diffraction spots, we measured the BFP image as a function of incident wavelength in the range of 1.3 – 2.25 eV. Plotted in Fig. 2(d) is the line profile in the y-direction $\boldsymbol{k} = (0, k_y)$, corresponding to a vertical cut through the center of the BFP images shown in Figs. 2(a)-(c). The bright diffraction spots observed with incident light at 1.65 eV (indicated A and B in Fig. 2(b)) now are shown in Fig. 2(d) to range from ~ 1.6—1.8 eV, appearing with wavenumbers with k$_y$ = ±0.34$k_d$, labeled A and B. The positions lie just beyond the light-line in air, indicated with the dotted red line. The additional diffraction spots with smaller wavenumbers show extended intensities with less pronounced spectral dependence and are likely due to the diffraction of light by the beam stop.

We assign the high intensity diffraction signals indicated A and B to a surface lattice resonance. The individual particles within the Sierpinski carpet structure exhibit a resonant interaction that occurs at a photon energy just after the disappearance of a diffraction maximum of the fractal leading to a high intensity surface wave on the air side. Energy in this excited mode can leak out into the glass substrate and can be detected in the BFP, in the same way that propagating surface plasmons on metal films can be observed using leakage radiation microscopy [21, 22]. The critical elements in our setup are excitation of a surface lattice resonance at the air side, and detection with an oil-immersion objective which extends the viewing angle. The former condition was validated by immersing the structure in refractive-index matching oil (silicone oil with refractive index 1.48) and repeating the measurements of the diffraction patterns. Spots A and B are then no longer present (see supplementary information). Under these conditions, the efficient coupling due to the constructive interference of the incident wave and scattered light from neighboring particles disappears. The condition for the surface lattice resonance is then shifted towards lower energies (longer wavelengths) but is obviously not detectable anymore in our scheme because a mismatch

## 5. NUMERICAL CALCULATIONS OF THE BACK-FOCAL PLANE

To confirm our experimental observations and explore the nature of the bright diffraction spots appearing beyond the light line, we calculate the far-field diffraction pattern of the Sierpinski carpet optical antenna at different incident wavelengths. We employ a finite-difference time-domain (FDTD) numerical method (Lumerical Solutions, Inc.) that includes the effect of a glass substrate as well as multipolar interactions between the rectangular monomers (see supplementary information for details). We obtain the diffraction pattern from the far-field transformation of the electric field present in the plane 10 nm below the sample in the glass substrate. The results presented in Fig. 2(e) confirm our experimental results that a very effective diffraction into wavevectors with $k_y = \pm 0.34 k_d$, occurs just below the light line in air.

The FDTD method, however, does not give a mechanistic view on or explanation of the observed efficient outcoupling of the surface lattice resonance. To obtain more insight we perform numerical calculations using the coupled dipole approximation [32-34]. This method has been applied successfully to the periodic case, where analytical results can be obtained [15] and in smaller systems of coupled plasmonic resonators such as plasmonic heptamers [10]. To perform the coupled dipole approximation we chose to simplify our system: we consider only the dipolar interaction between monomers and do not include the effect of the glass substrate. In short in the coupled dipole approximation, we calculate the induced dipole moment $\boldsymbol{P}_m = (P_x, P_y, P_z)$ of each monomer under the x-polarized plane wave excitation. Where a monomer is modelled as a spherical particle of the same volume as in the experiment (1.6 10$^6$ nm$^3$). The coupled dipole equation is:

$$\boldsymbol{P}_m = \alpha \left[ \boldsymbol{E}_{ext,m} + \sum_{n \neq m} \overline{\boldsymbol{M}}_{m,n} \boldsymbol{P}_n \right] \quad (1)$$

where $\boldsymbol{P}_m$ is the dipole moment of m$^{th}$ monomer, $\overline{\boldsymbol{M}}_{m,n}$ is the interaction matrix between a receiving dipole at $\boldsymbol{r}_m$ and a radiating dipole at $\boldsymbol{r}_n$. Here we use the Green's function in air and thus intentionally neglect the substrate in the calculation. According to the fundamental principle of Fourier optics, the far-field electric field is entirely defined by the Fourier spectrum of the electric field in the object plane [5]. Therefore, we calculate the Fourier spectrum of the dipole moment $P_x$:

$$\tilde{P}_x(k_x, k_y) = |\int P_x(x,y) e^{-i(k_x x + k_y y)} dx dy| \quad (2)$$

We may neglect components $P_y$ and $P_z$ here because $|P_x| \gg |P_y|, |P_z|$ under the x-polarized plane wave excitation. Panel (f) in Fig 2 shows $\tilde{P}_x$ along the y-direction calculated by the coupled dipole approximation. Note that the image is not the far-field radiation pattern as the CDA as we implemented it ignores the glass substrate; it does provides a picture of the regions in k-space that are excited by the incident plane wave in the fractal structure in air. In this simulation, also the central region (blocked in the experiment and ignored in the FDTD simulation) is shown. The three panels (d), (e) and (f) of Fig. 2 all show the same distinctive features: vertical lines and bright regions marked by intense signals A and B. The vertical lines arise from the non-periodic analogue of the familiar diffraction grating formula: $\boldsymbol{k}_{dif} = \boldsymbol{k}_{inc} + \boldsymbol{G}_{frac}$. Given that the incident wave is perpendicular ($\boldsymbol{k}_{inc} = \boldsymbol{0}$) the parallel wavevector component of the diffracted light $\boldsymbol{k}_{dif}$ is equal to the lattice vector of the fractal $\boldsymbol{G}_{frac}$, which is independent on the photon energy. In the finite and non-periodic fractal structure $\boldsymbol{G}_{frac}$ can be replaced by the magnitude of the Fourier transform of the structure, therefore vertical lines appear, corresponding to the large magnitude Fourier components of the Sierpinski carpet optical antenna (see also Fig 1c).

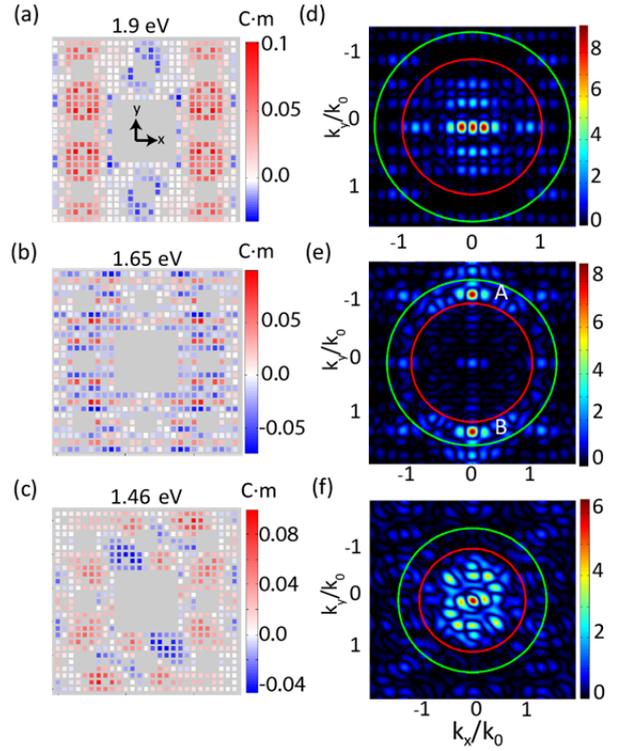

Fig. 3 Panels (a) - (c) represent the real part of $P_x$ of the most efficiently driven eigenmode with light incident at 1.9, 1.65 and 1.46 eV. Panels (d) - (f) show the calculated Fourier amplitude spectra of $P_x$ in (a) - (c), respectively, where the red and green circles represent the wavevector of the light line in air and in the glass substrate, respectively. $k_0$ is the wavevector of incident light in air.

The striking features A, B in the diffraction of the Sierpinski fractal structure are very well described by the highly simplified coupled dipole model. Note that switching off the particle interactions by making $\bar{M}_{m,n} = 0$ removes the high intensity feature in the calculated diffraction pattern (results not shown). The observed resonant feature at A, B in Fig 2 d,e,f therefore corresponds to in-plane diffracted light on the air side of the sample that leaks out of a collective mode that is analog to the surface lattice resonance in periodic structures.

## 6. EIGENMODE ANALYSIS

To further disentangle the complex interactions in the fractal structure, we extract the dominant eigenmode under normal excitation by the eigen-decomposition method [35, 36]. Here, the dominant eigenmode is defined as the eigenmode with largest weight among all eigenmodes. Figs. 3(a) – 3(c) and 3(d) - 3(f) show the dipole moment distribution $P_x$ and their corresponding Fourier amplitude of the dominant eigenmodes of the fractal at excitation photon energy 1.9, 1.65 and 1.46 eV, respectively. We find that for the mode at 1.65 eV, most of the wavevectors are located outside the light line of the incident light in air (the red circle) as shown in Fig. 3(e), indicating it is sub-radiant. This is also observable in the dipole moment distribution in Fig. 3(b) in which the periodicity of the positive polarized rows (red) matches the length of 3 monomers enabling only a weak coupling to the far-field on the air side. On the contrary, the dominant eigenmodes at 1.9 and 1.46 eV shown in Figs. 3(d) and 3(f) respectively, have large amplitudes inside the light cone and are thus bright (radiant). We recognize the intense spots A and B in Fig. 3(e), which are associated with the surface lattice resonance. We conclude here separately that the signals A and B in Fig. 2 arise from the resonance of the dark eigenmode shown in Fig. 3(b). It has to be noted, however, that out of the 1536 (partially degenerate) eigenmodes of the structure, about 250 contribute with some significant amplitude and the dominant eigenmode has an amplitude contribution of about 7% (see supplementary information).

## 7. FANO PROFILES

An important characteristic of a subradiant eigenmode such as a surface lattice resonance is that it can generate a Fano profile in the extinction or scattering spectrum of plasmonic nanostructures [7, 23, 15]. In general, Fano profiles arise from the destructive interference between a resonant mode and a continuum. In plasmonic structures this can arise in different forms, for instance in (1) the interference between a dark and a bright eigenmode or (2) the interference between a plasmonic resonance and the directly transmitted light. Unfortunately the latter is difficult to measure in a relatively small structure as described in this article because the scattering is of much lower magnitude than the direct transmission. We therefore aimed to determine if a Fano-like interference between dark and

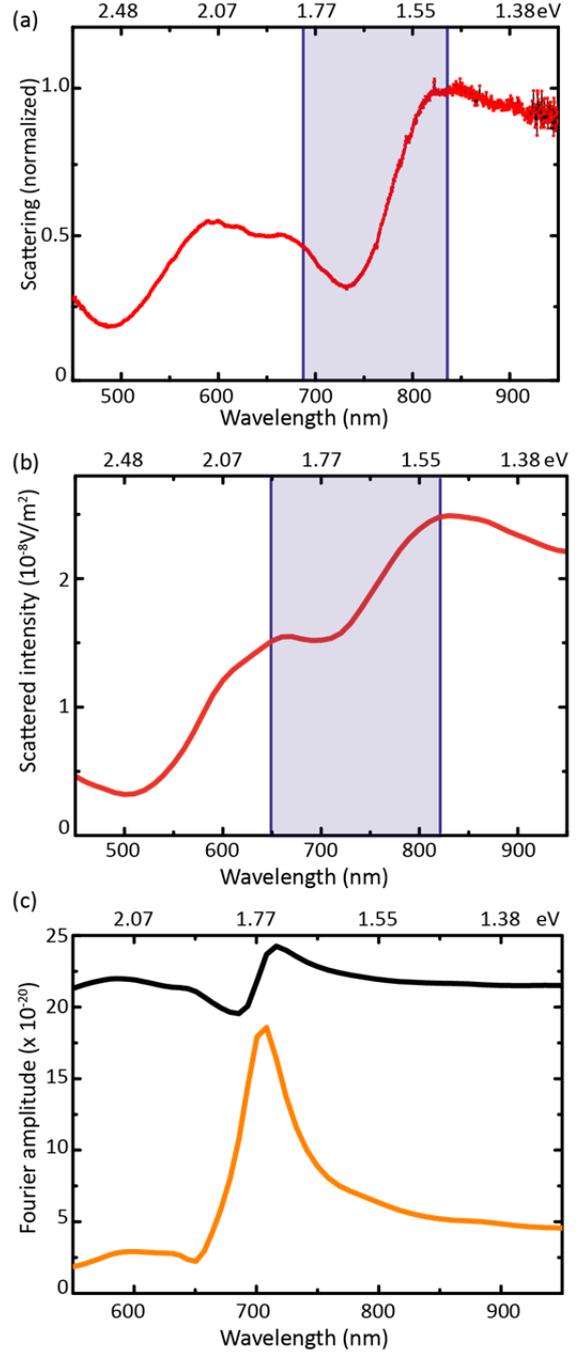

Fig. 4 (a) The measured scattering spectrum of the Sierpinski carpet optical antenna with x-polarized incident light. The shaded zone corresponds to the region of intense signal A, B in Fig. 2(d). (b) The scattering cross section calculated by the FDTD method. The shaded region corresponds to the photon energy range of where the surface lattice resonance was found in FDTD (Fig 2e). (c) Numerical results from the coupled dipole method. In black the light scattered in forward direction $\boldsymbol{k} = (0,0)$, in orange the light scattered in $\boldsymbol{k} = (0, 2\pi/3\mathrm{d})$.

bright modes can be detected in scattering, despite the fact that more than 1500 modes are present in the structure.

Fig. 4(a) shows the measured scattering spectrum of the Sierpinski carpet optical antenna, obtained by white-light dark field microscopy [29]. This measurement was performed using an annular beam of white light incident from the glass side on the sample with polar angles ranging from 20° - 50°. The scattered light in air was collected using a low NA objective (see supplementary information for details of the setup). The scattering spectrum features a significant dip around 730 nm, which coincides with the spectral region where the intense signals A and B appeared in the back focal plane images (shaded region in Fig. 4(a)).

In Fig. 4(b) the scattering of the structure under similar conditions as the experiment was calculated by FDTD. Shown is the light scattered into the glass in the 20° - 50° cone when light is incident normal to the surface which via reciprocity should yield a similar spectrum. The shaded region corresponds to the region where the surface lattice resonance is present in FDTD calculations. The scattering spectrum of Fig. 4(b) indicates a small dip around 720 nm, which is close to the measured result in Fig. 4(a). Though both experiment and FDTD show an indication of a Fano-like profile, the correspondence is not of high enough quality to merit a strong conclusion.

We therefore invoke once more the results from the coupled dipole approximation to determine if the drop in scattering intensity around 730 nm (Fig. 4(a)) could indeed be a Fano dip. Fig. 4(c) shows line profiles of the Fourier spectra of $P_x$ calculated by the coupled dipole approximation that are taken at the position of the black and orange arrows drawn in Fig. 2(f) at $\boldsymbol{k} = (0,0)$ and $\boldsymbol{k} = \left(0, \frac{2\pi}{3d}\right)$ respectively. The forward scattering (black curve) clearly has a Fano line shape with a minimum at around 690 nm and maximum at around 710 nm. It corresponds very well to the spectral position of the surface lattice resonance visible as the orange curve. We thus conclude that the Fano line shape visible in the forward direction arises from the π phase jump of the dominant eigenmode, which interferes respectively destructively and constructively with bright eigenmodes of the fractal structure. The destructive interference occurs within the red circle of BFP images, which causes the Fano dip in the scattering spectrum. This interpretation agrees with Frimmer et al. [10], who state that the Fano dip requires an overlap of the far-field radiation pattern of a dark and bright eigenmode.

We believe the same type of interference between modes is observed in the experiments here, however, higher-order multipolar interactions and the leakage of the mode into the glass substrate lead to overall broadening and deviations from the simplified result obtained from the coupled dipole approximation. Also due to the complex mode structure of the fractal antenna and broad spectral features of plasmonics in the visible, it is challenging to show exactly which or how many modes are needed to generate the scattering spectrum of the fractal structure.

## 8. DISCUSSION

The existence of a surface lattice resonance in a fractal arrangement of plasmonic resonators is clearly demonstrated; however, several intriguing side issues remain. Firstly the surface lattice resonance occurs when the light-line crosses the 2π/3d maximum in Fourier amplitude of the fractal structure. Though the amplitude of this component is certainly among the stronger peaks in the Fourier spectrum, it is unclear why conditions for a surface lattice resonance are not met at other crossings. A possible solution of this lies in a detailed study of the eigenmodes that are extracted based on the CDA, though the large number of modes makes this a very complex puzzle. We tentatively attribute the surface lattice resonance to the fractal arrangement of our Sierpinski carpet optical antenna, which renders an efficient subradiant mode with high eigenpolarizability. Closely related, the second issue is how such a clear Fano resonance emerges from the coupled dipole approximation whereas a large number of modes is collectively responsible for the response of the structure. Further research is needed to unravel the complex behavior of this fractal plasmonic structure.

## 9. CONCLUSION

In conclusion, we demonstrate a surface lattice resonance exists in the Sierpinski carpet optical antenna, which is triggered analogous to the way these resonances occur in periodic particle arrays: the coupling of plasmonic resonators via in-plane diffracted waves. These resonances were visualized by performing the experiment with a structure on a non-index matched geometry using leakage radiation microscopy, albeit at the cost of higher radiation losses. The surface lattice resonance emerges in the analysis using the eigen-decomposition method as a subradiant dominant eigenmode with a dominant large k-space contribution that can be assigned to in-plane diffracted waves. The findings agree with surface lattice resonances in aperiodic structures [17, 19, 20] indicating that they exist in many other non-periodic lattices. The leakage radiation method to visualize the surface lattice resonance we used can be used in other non-periodic and periodic structures. Our findings can be used in the design of nanophotonic devices and plasmonic sensors [2-4, 37] which exploit aperiodic or fractal plasmonic metal nanoparticle arrays.

**Funding sources and acknowledgments**

We thank Allard Mosk for fruitful discussions. This research is financially supported by Nederlandse Organisatie voor Wetenschappelijk Onderzoek (NWO). The research of JLH is supported by an NWO-Vici grant; JCP acknowledges support from Stichting Technische

Wetenschappen (STW) under the nanoscopy program (project nr.12149).## References

1. C. F. Bohren and D. R. Huffman, *Absorption and Scattering of Light by Small Particles* (John Wily & Sons, 1983).
2. N. Liu, M. L. Tang, M. Hentschel, H. Giessen, and a P. Alivisatos, "Nanoantenna-enhanced gas sensing in a single tailored nanofocus.," Nat. Mater. **10**, 631–636 (2011).
3. P. Offermans, M. C. Schaafsma, S. R. K. Rodriguez, Y. Zhang, M. Crego-Calama, S. H. Brongersma, and J. Gómez Rivas, "Universal scaling of the figure of merit of plasmonic sensors.," ACS Nano **5**, 5151–7 (2011).
4. G. Lozano, D. J. Louwers, S. R. Rodríguez, S. Murai, O. T. Jansen, M. a Verschuuren, and J. Gómez Rivas, "Plasmonics for solid-state lighting: enhanced excitation and directional emission of highly efficient light sources," Light Sci. Appl. **2**, e66 (2013).
5. L. Novotny and B. Hecht, *Principles of Nano-Optics* (Cambridge University Press, 2007).
6. E. Prodan, C. Radloff, N. J. Halas, and P. Nordlander, "A hybridization model for the plasmon response of complex nanostructures.," Science **302**, 419–22 (2003).
7. N. I. Zheludev, S. A. Maier, N. J. Halas, P. Nordlander, H. Giessen, and C. T. Chong, "The Fano resonance in plasmonic nanostructures and metamaterials," Nat. Mater. **9**, 707–715 (2010).
8. D. E. Gómez, Z. Q. Teo, M. Altissimo, T. J. Davis, S. Earl, and a Roberts, "The dark side of plasmonics.," Nano Lett. **13**, 3722–8 (2013).
9. F. Shafiei, F. Monticone, K. Q. Le, X.-X. Liu, T. Hartsfield, A. Alù, and X. Li, "A subwavelength plasmonic metamolecule exhibiting magnetic-based optical Fano resonance.," Nat. Nanotechnol. **8**, 95–9 (2013).
10. M. Frimmer, T. Coenen, and a. F. Koenderink, "Signature of a Fano Resonance in a Plasmonic Metamolecule's Local Density of Optical States," Phys. Rev. Lett. **108**, 077404 (2012).
11. E. M. Hicks, S. Zou, G. C. Schatz, K. G. Spears, R. P. Van Duyne, L. Gunnarsson, T. Rindzevicius, B. Kasemo, and M. Ka, "Controlling Plasmon Line Shapes through Diffractive Coupling in Linear Arrays of Cylindrical Nanoparticles Fabricated by Electron Beam Lithography," Nano Lett. **5**, 1065–1070 (2005).
12. F. J. G. De Abajo, "Colloquium: Light scattering by particle and hole arrays," Rev. Mod. Phys. **79**, 1267–1290 (2007).
13. Y. Chu, E. Schonbrun, T. Yang, and K. B. Crozier, "Experimental observation of narrow surface plasmon resonances in gold nanoparticle arrays," Appl. Phys. Lett. **93**, 181108 (2008).
14. V. G. Kravets, F. Schedin, and a. N. Grigorenko, "Extremely narrow plasmon resonances based on diffraction coupling of localized plasmons in arrays of metallic nanoparticles," Phys. Rev. Lett. **101**, 087403 (2008).
15. B. Auguié and W. Barnes, "Collective Resonances in Gold Nanoparticle Arrays," Phys. Rev. Lett. **101**, 143902 (2008).
16. G. Vecchi, V. Giannini, and J. Gómez Rivas, "Shaping the Fluorescent Emission by Lattice Resonances in Plasmonic Crystals of Nanoantennas," Phys. Rev. Lett. **102**, 146807 (2009).
17. L. Dal Negro and S. V. Boriskina, "Deterministic aperiodic nanostructures for photonics and plasmonics applications," Laser Photon. Rev. **6**, 178–218 (2012).
18. C. Forestiere, G. F. Walsh, G. Miano, and L. Dal Negro, "Nanoplasmonics of prime number arrays.," Opt. Express **17**, 24288–303 (2009).
19. C. Bauer, G. Kobiela, and H. Giessen, "2D quasiperiodic plasmonic crystals.," Sci. Rep. **2**, 681 (2012).
20. S. M. Lubin, A. J. Hryn, M. D. Huntington, C. J. Engel, and T. W. Odom, "Quasiperiodic moiré plasmonic crystals.," ACS Nano **7**, 11035–42 (2013).
21. B. Hecht, H. Bielefeldt, L. Novotny, Y. Inouye, and D. Pohl, "Local Excitation, Scattering, and Interference of Surface Plasmons," Phys. Rev. Lett. **77**, 1889–1892 (1996).
22. A. Drezet, A. Hohenau, A. L. Stepanov, H. Ditlbacher, B. Steinberger, N. Galler, F. R. Aussenegg, A. Leitner, and J. R. Krenn, "How to erase surface plasmon fringes," Appl. Phys. Lett. **89**, 091117 (2006).
23. M. Hentschel, M. Saliba, R. Vogelgesang, H. Giessen, a P. Alivisatos, and N. Liu, "Transition from isolated to collective modes in plasmonic oligomers.," Nano Lett. **10**, 2721–6 (2010).
24. M. V Berry, "Diffractals," J. Phys. A. Math. Gen. **12**, 781–797 (1979).
25. C. Allain and M. Cloitre, "Optical diffraction on fractals," Phys. Rev. B **33**, 3566–3569 (1986).
26. M. Segev, M. Soljačić, and J. M. Dudley, "Fractal optics and beyond," Nat. Photonics **6**, (2012).
27. G. Volpe, G. Volpe, and R. Quidant, "Fractal plasmonics: subdiffraction focusing and broadband spectral response by a Sierpinski nanocarpet.," Opt. Express **19**, 3612–8 (2011).
28. S. Sederberg and A. Y. Elezzabi, "Sierpiński fractal plasmonic antenna: a fractal abstraction of the plasmonic bowtie antenna," Opt. Express **19**, 10456–61 (2011).
7

29. T. L. Chen, D. J. Dikken, J. C. Prangsma, F. Segerink, and J. L. Herek, "Characterization of Sierpinski carpet optical antenna at visible and near-infrared wavelengths," New J. Phys. **16**, 093024 (2014).
30. R. G. Hohlfeld and N. Cohen, "Self-similarity and the geometric requirements for frequency independence in antennae," Fractals **07**, 79–84 (1999).
31. J.-S. Huang, V. Callegari, P. Geisler, C. Brüning, J. Kern, J. C. Prangsma, X. Wu, T. Feichtner, J. Ziegler, P. Weinmann, M. Kamp, A. Forchel, P. Biagioni, U. Sennhauser, and B. Hecht, "Atomically flat single-crystalline gold nanostructures for plasmonic nanocircuitry.," Nat. Commun. **1**, 150 (2010).
32. W. Weber and G. Ford, "Propagation of optical excitations by dipolar interactions in metal nanoparticle chains," Phys. Rev. B **70**, 125429 (2004).
33. B. T. Draine and P. J. Flatau, "Discrete-dipole approximation for scattering calculations," J. Opt. Soc. Am. A **11**, 1491 (1994).
34. V. L. Y. Loke, M. Pinar Mengüç, and T. a. Nieminen, "Discrete-dipole approximation with surface interaction: Computational toolbox for MATLAB," J. Quant. Spectrosc. Radiat. Transf. **112**, 1711–1725 (2011).
35. V. A. Markel, "Antisymmetrical optical states," J. Opt. Soc. Am. B **12**, 1783 (1995).
36. K. H. Fung and C. T. Chan, "Plasmonic modes in periodic metal nanoparticle chains: a direct dynamic eigenmode analysis," Opt. Lett. **32**, 973 (2007).
37. J. N. Anker, W. P. Hall, O. Lyandres, N. C. Shah, J. Zhao, and R. P. Van Duyne, "Biosensing with plasmonic nanosensors," **7**, 8–10 (2008).
8

# Supplementary materials: Observation of a surface lattice resonance in a fractal arrangement of gold nanoparticles

Ting Lee Chen[1], Jord C. Prangsma[1,2], Frans Segerink[1], Dirk Jan Dikken[1], and Jennifer L. Herek[1]

[1]Optical Sciences Group, MESA+ Institute for Nanotechnology, University of Twente, 7500 AE Enschede, The Netherlands

[2]Nanobiophysics, MESA+ Institute for Nanotechnology, University of Twente, 7500 AE Enschede, The Netherlands

## Contents

### Sample fabrication

To fabricate our samples, we used focused ion beam (FIB) milling (FEI Nova 600 dual beam) to carve nanostructures directly out of single-crystalline gold flakes. We deposited single crystalline gold flakes onto a glass substrate with a 100 nm ITO coating. The flake was pre-thinned to the thickness of ~40 nm with FIB milling. In the FIB milling process, the acceleration voltage was 30 kV and the Ga-ion current was 1.5 pA, structures were milled using 30 passes of the ion beam.

### The measurement setup and principle - leakage radiation microscopy with wavelength tunable light source

Fig. S1 shows the schematic of the leakage radiation microscopy. The photon energy of the excitation light source can be tuned from 1.3 to 2.25 eV (550 – 950 nm) with spectral bandwidth of 2 nm by sending the broadband white light (Fianium SC400-4) into a monochromator (Acton SP2100-i). The excitation light is focused by a 0.3 NA objective (UPLFLN 10X2, Olympus) the focus is adjusted to cover the whole sample. The diffracted light is collected by a 1.4 N.A. oil-immersion objective (UPLSAPO 100XO, Olympus). The back focal plane (BFP) is imaged by lenses I, II and III in Fig. 1(b) onto the CCD. A beam block (BB) is used for blocking the incident light. With lenses I, II and IV, we can alternatively display the image at the object plane (OP) onto the CCD. With the polarizer and analyzer (GT10, Thorlabs) inserted as shown in Fig. S1, we can choose the BFP image with x and y polarization.

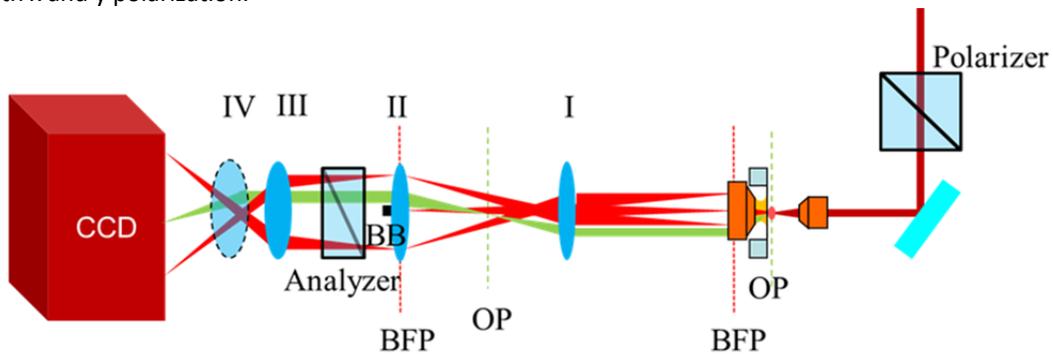

*Fig. S1.* Schematic of the leakage radiation microscopy. BFP: back focal plane; BB: beam blocker; OP: object plane. The red beams represent the imaging of the back-focal plane, the green beam represents the light from the object plane. The zero order diffracted light and incident light are blocked by the beam block.

### The measurement of the scattering spectrum of the Sierpinski carpet optical antenna

To measure far-field scattering properties of the structures, we use a home-built dark-field spectrometer setup as shown schematically in Fig. S2. (a). White light from a Xenon Arc lamp (Oriel 71213, Newport) is sent into a microscope (IX71, Olympus) and focused onto samples with a 1.4 N.A. oil-immersion objective (UPLSAPO 100XO, Olympus). To obtain a dark-field illumination, the central part of light beam was blocked. The scattered light was collected with a 0.3 N.A. objective (UPLFLN 10X2, Olympus), and subsequently focused onto a 50 μm diameter pinhole before the spectrometer (AvaSpec-3648-USB2, Avantes). The effective area on the sample plane from which light was collected was estimated to be a 5 μm diameter circle, hence completely encompassing the full nanostructure. Spectra were typically acquired in 5 seconds to allow the detector to accumulate enough signal. The scattering spectra are normalized to the system response retrieved by removing the beam block (bright field illumination) to remove the inherent wavelength dependence of white light source.

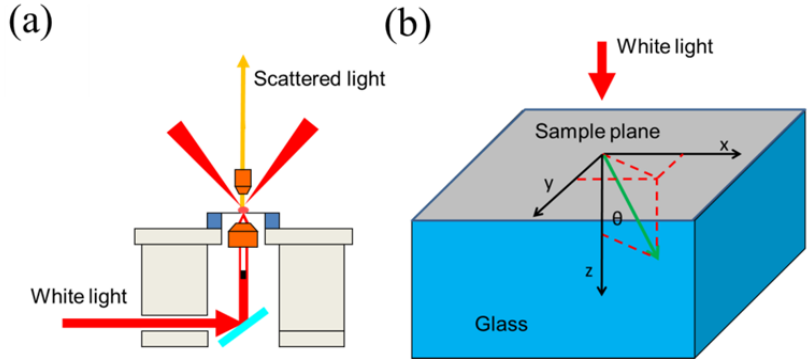

***Fig. S2.*** *A schematic of white-light dark field microscopy is shown in (a). Note the inclusion of a beam block for annular illumination. In (b), the collection angle used in numerical simulation of the scattering spectra is illustrated. The polar angle ϑ for the collection of scattered light is from 20º to 50º. In this configuration, the 0$^{th}$ order diffracted light is excluded.*

**Measurement with sample immersed in refractive index-matching oil**

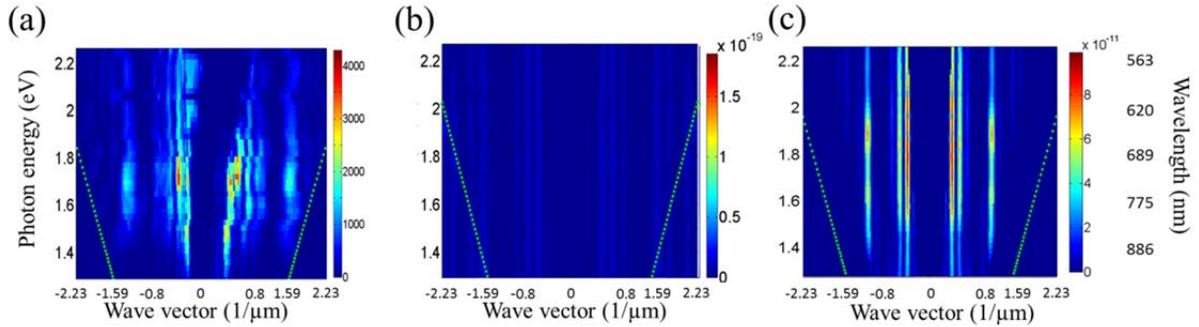

***Fig. S3.*** *(a), similar to Fig. 2(a), is the line profiles as measured with the Sierpinski carpet optical antenna immersed in the refractive index-matching oil. (b) and (c), similar to Fig. 2(b) and Fig. 2(c), are the line profiles calculated with the Sierpinski carpet optical antenna immersed in the oil by the coupled dipole approximation and FDTD method, respectively. The green dot lines in each picture represent the wave vector of light line in the glass substrate.*

**The Finite-difference time domain (FDTD) simulation on diffraction patterns of the Sierpinski carpet optical antenna**

We used FDTD (Lumerical Solutions, Inc.) to calculate the scattering spectra and far-field patterns. For the far-field patterns the electric field at a plane of 10 nm above (in the air) and below (in the glass substrate) the Sierpinski carpet optical antenna was monitored. The schematic drawings of the position of plane are shown in the pictures above (a) and (g) of Fig. S4. The structure geometry is modeled as an arrangement of gold rectangles with dimension of 200×200×40 nm$^3$ (W×L×H) on a glass substrate. The dielectric constants gold and glass are modeled as fits on experimental data from [S1] and [S2], respectively. The mesh refinement is "conformal variant 1" and the minimum mesh size is 4 nm. The total-field scattered-field (TFSF) plane wave source with normal incidence on the gold nanostructures is used as excitation source.

Fig. S4(a) – S4(c) show the calculated electric field at the plane of 10 nm below the interface between the sample and glass substrate with incident light wavelength at 1.9, 1.65 and 1.46 eV, respectively. The dielectric constant of gold and glass substrate in the FDTD method is from the experimental data Ref.[S1] and [S2], respectively. Figs. S4(d) – S4(f) are the far field projection of the electric fields shown in Figs. S4(a) – S4(c), respectively. Figs. S4(g) – S4(i) show the calculated electric field at the plane of 10 nm above the sample (in the air) with incident wavelength at 1.9, 1.65 and 1.46 eV, respectively. Figs. S4(j) – S4(l) are the far field projection of the electric fields shown in Figs. S4(g) – S4(i), respectively. Though Figs. S4(a) – S4(c) and Figs. S4(g) – S4(i) are near-field electric field, their far field projection are different: the far field projection of the electric field in the glass in Figs. S4(d) – S4(f) contains wave vectors which can't be seen in Figs. S4(j) – S4(l). These wave vectors are corresponding to the evanescent waves in the air.

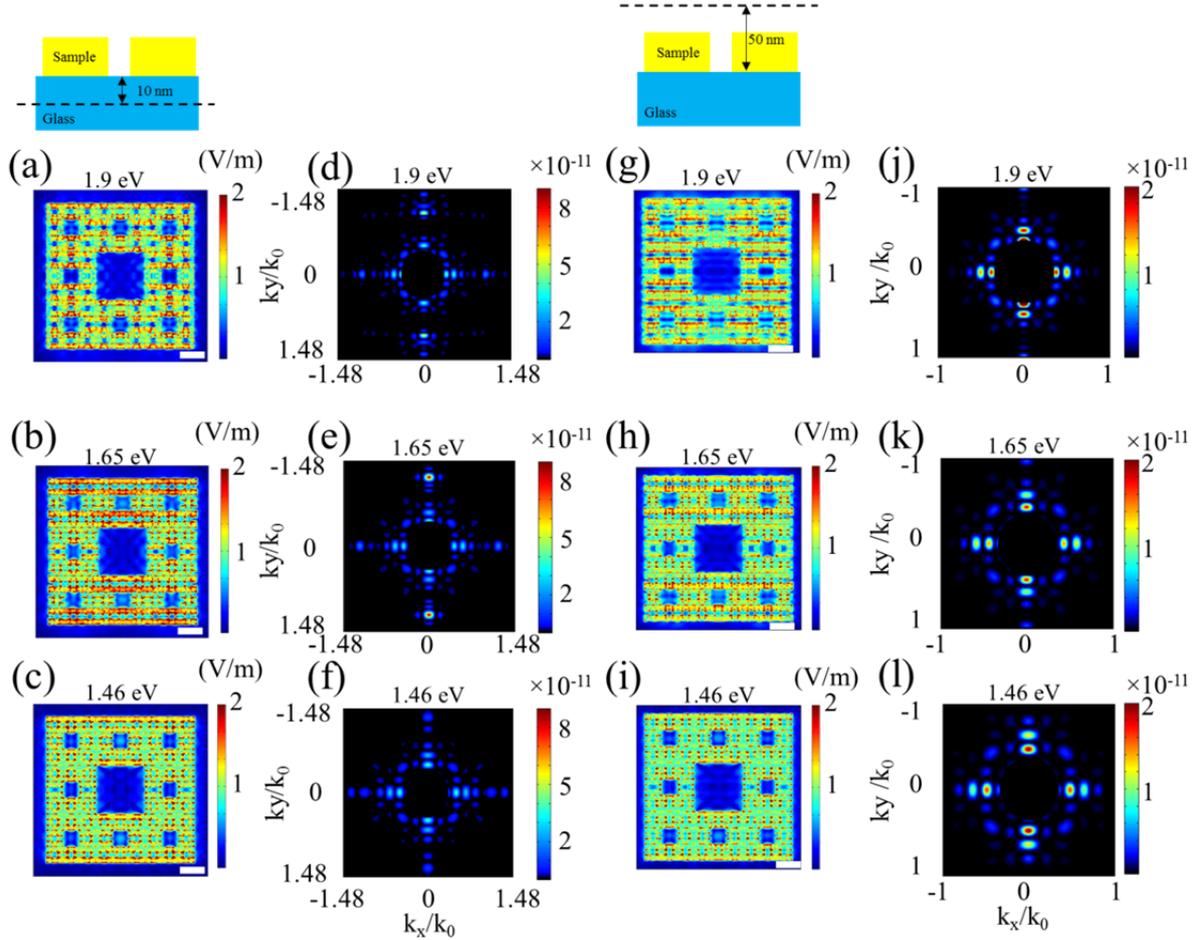

*Fig. S4.* *(a) - (c) are the electric field at the plane of 10 nm below the interface between the sample and glass substrate with the incident light wavelength at 1.9, 1.65 and 1.46 eV. (d) - (f) are the far field projection of electric fields (a) - (c), respectively. (g) - (i) are the electric field at the plane of 50 nm above the interface with the incident light wavelength at 1.9, 1.65 and 1.46 eV. (j) - (l) are the far field projection of electric fields (g) - (i), respectively. The unit of the color bar for (a) - (c) and (g) - (i) is V/m and (V/m)$^2$ for (d) - (f) and (j) - (l). Note that the region of $0^{th}$ order diffracted light in (d) - (f) and (j) - (l) is set to 0 for a better comparison with the BFP images in Fig. 2 (a) - 2(c).*

**The Finite-difference time domain (FDTD) simulation of the scattering spectrum of the Sierpinski carpet**

Numerical results were obtained using finite difference time domain (FDTD, Lumerical Solutions Inc.) calculations. The settings of FDTD method is the same as the calculation for the diffraction patterns of the Sierpisnki carpet optical antenna. To simulate the experiment conditions, we use a reversed configuration shown in Fig. S2(b) : the white light incident normally on the sample from the air side, and the collection polar angle θ of the scattered light is set to be 20º to 50º in the side of glass substrate. We calculate the near-field electric field at the plane of 10 nm below the interface between the sample and glass substrate, then we use the "near to far field projections" in the Lumerical FDTD method to get the far-field radiation pattern. From the far-field radiation pattern at different wavelengths, we can sum the total far-field radiation power within the collection angle θ = 20º - 50º at each wavelengths to get the scattering spectrum. This does not include the contribution of the power from dominate sub-radiant eigenmode. Fig. S5 shows an example of the far field radiation pattern at wavelength 1.77 eV. The range of collection angle θ for Fig. S9 is 20º - 50º.

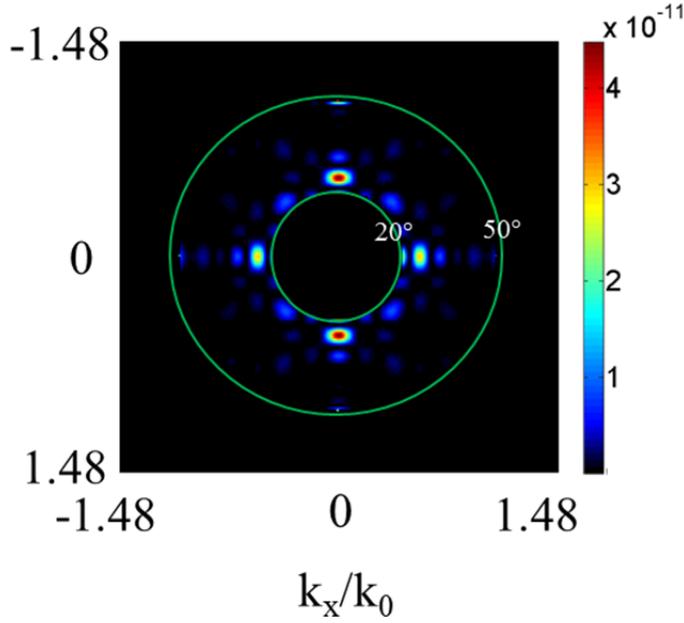

**Fig. S5.** *The far field intensity at 1.77 eV calculated with FDTD. The range of collection polar angle is 20º - 50º.*

### The coupled dipole approximation

The coupled dipole equation is:
$$P_m = \alpha\left[E_{ext,m} + \sum_{n \neq m} M_{m,n} P_n\right] \quad (S1)$$
, where $P_m$ is the dipole moment of $m^{th}$ monomer, $M_{m,n}$ is the tensor that represents the interaction between a receiving dipole at $\vec{r}_m$ and a radiating dipole at $\vec{r}_n$.

$M_{m,n}$ is defined as:
$$M_{m,n}(\vec{r} \equiv (\vec{r}_m - \vec{r}_n)) = k_0^3 \left[B(k_0|\vec{r}|)\delta_{ij} + C(k_0|\vec{r}|)\frac{r_i r_j}{|\vec{r}|^2}\right] \quad (S2)$$
$$B(x) = (x^{-1} + ix^{-2} - x^{-3})e^{ix} \quad (S3)$$
$$C(x) = (-x^{-1} - 3ix^{-2} + 3x^{-3})e^{ix}, \quad (S4)$$

where $k_0$ is wave vector of the incident light. The polarizability α here is the dynamic dipole polarizability:
$$\alpha(\omega) = i\frac{3c^3}{2\omega^3} a_1(\omega), \quad (S5)$$
where
$$a_1(\omega) = \left.\frac{\varepsilon(\omega)j_1(k_s r)\partial_r[rj_1(k_0 r)] - \varepsilon_h j_1(k_0 r)\partial_r[rj_1(k_s r)]}{\varepsilon(\omega)j_1(k_s r)\partial_r[rh_1^{(1)}(k_0 r)] - \varepsilon_h h_1^{(1)}(k_0 r)\partial_r[rj_1(k_s r)]}\right|_{r=r_0} \quad (S6)$$

is the Mie's coefficient for L=1 electric term, $j_1(x)$ is the spherical Bessel function of the first kind, $h_1^{(1)}(x)$ is the spherical Bessel function of the third kind (or spherical Hankel function of the first kind), and $k_s = \omega(\varepsilon(\omega))^{1/2}/c$, $\varepsilon(\omega)$ is the dielectric function from the interpolation of experimental data [S2] and c is the speed of light in the air. The effective radius of the monomer of the Sierpinski carpet optical antenna was chosen 72.5 nm in the calculation of the coupled dipole approximation, which was obtained by $r_0 = (3V_c/4\pi)^{1/3}$, where $V_c$ is the volume of a monomer.

### Calculation of the dominant eigenmode with the eigen-decomposition method

We diagonalize the interaction matrix $M_{m,n}$ in the coupled dipole approximation to get the 1536 (512×3) eigenvectors $|P_i>$, i = 1, 2,…, 1536. The dipole moment P of the optical antenna excited by a x-polarized plane wave can be written as the linear combination of $|P_i>$ : $|P> = \sum_i c_i |P_i>$, where $|P>$ is the complex column vector of dipole moment and $c_i$ is the complex, too. The dominant eigenmode for $|P>$ is defined as the individual eigenmode $|P_i>$ having maximum $|c_i|$ among all eigenmodes. To obtain the coefficient $|c_i|$, we multiply $<P_j|$ with $|P> : <P_j|P> = \sum_i c_i <P_j|P_i>$. The indices i, j = 1, 2,…, 1536. Note that $<P_j|P_i> \neq \delta_{ij}$, where $\delta_{ij} = 1$ for i = j, $\delta_{ij} = 0$ for i ≠ j, because eigenvectors $|P_i>$, i = 1, 2,…, 1536, are not an orthogonal set [S3]. Therefore, we solve 1536 linear equations to determine $c_i$.

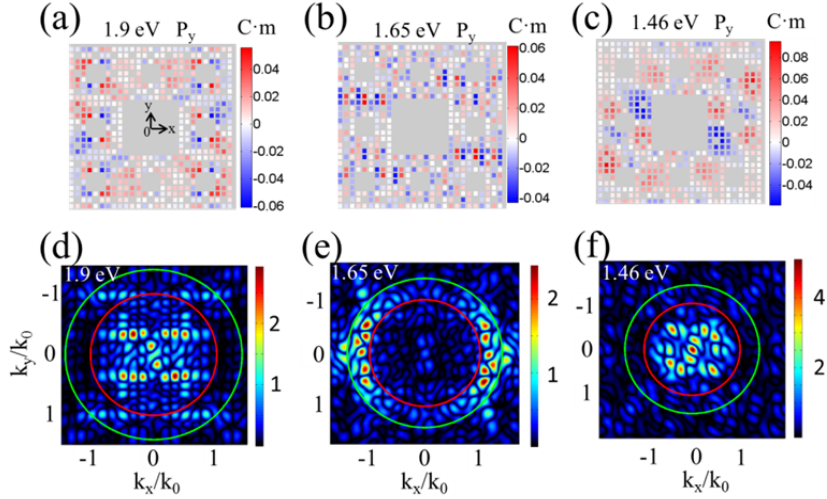

***Fig. S6.*** *(a) - (c) are the real part of $P_y$ of the dominate eigenmode at excitation wavelength 1.9, 1.65 and 1.46 eV. (d) - (f) show the Fourier spectra of $P_y$ at 1.9, 1.65 and 1.46 eV. In (d) - (f) the red and green circle represent wave vector of light line in the air and glass substrate respectively. $k_0$ is the wave vector of incident light in the air.*

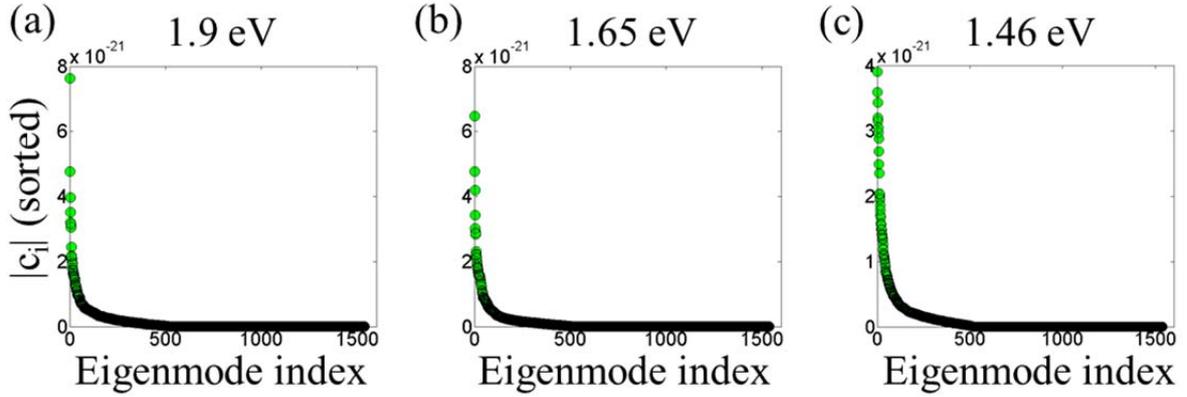

***Fig. S7.*** *The sorted $|c_i|$ of each eigenmode of the Sierpinski carpet optical antenna for (a) 1.9 eV, (b) 1.65 eV, (c) 1.46 eV. For (a), (c) the dominate eigenmode ( the eigenmode having maximum $|c_i|$) are bright and for 1.65 eV is dark as shown in Fig. 4 and Fig. S6.*